\documentclass[
reprint,
twocolumn,
showpacs,
 amsmath,amssymb,
 aps,
 prl,
]{revtex4}

\usepackage{amsthm}

\usepackage[dvipsnames]{xcolor}

\usepackage{graphicx}
\usepackage{dcolumn}
\usepackage{bm}
\usepackage{hyperref}
\usepackage[utf8]{inputenc}
\usepackage[english]{babel}
\usepackage[notref,notcite,final]{showkeys}

\RequirePackage[normalem]{ulem}
\RequirePackage{color}

\begin{document}
\title{Stochastic Laplacian growth}
\author{Oleg Alekseev}
\author{Mark Mineev-Weinstein}

\affiliation{%
International Institute of Physics,\\ Federal University of Rio Grande do Norte, 59078-970, Natal, Brazil
}%

\begin{abstract}
A point source on a plane constantly emits particles which rapidly diffuse and then stick to a growing cluster.  The growth probability of a cluster is presented as a sum over all possible scenarios leading to the same final shape.  The classical point for the action, defined as a minus logarithm of the growth probability, describes the most probable scenario and reproduces the Laplacian growth equation, which embraces numerous fundamental free boundary dynamics in non-equilibrium physics.  For non-classical scenarios we introduce virtual point sources, in which presence the action becomes the Kullback-Leibler entropy.  Strikingly, this entropy is shown to be the sum of electrostatic energies of layers grown per elementary time unit. Hence the growth probability of the presented non-equilibrium process obeys the Gibbs-Boltzmann statistics, which, as a rule, is not applied out from equilibrium.   Each layer's probability is expressed as a product of simple factors in an auxiliary complex plane after a properly chosen conformal map. The action at this plane is a sum of Robin functions, which solve the Liouville equation.  At the end we establish connections of our theory with the tau-function of the integrable Toda hierarchy and with the Liouville theory for non-critical quantum strings.
\end{abstract}
\maketitle

{\it The goal of this work} is to unify two fundamental highly non-equilibrium processes, {\it Laplacian growth} (LG)~\cite{Pelce,MWZ,Gillow,MPT, GTV}, which is deterministic interface dynamics, and {\it diffusion-limited aggregation} (DLA)~\cite{WS81} -- a discrete universal stochastic fractal growth. These remarkable processes have a lot in common and were suspected to be deeply related~\cite{D, LG-review, Arneodo, PRL94,Hastings-Levitov}.

{\it Laplacian growth} raised enormous interest in {\it physics} because of (i) its impressively wide applicability ranging from solidification and oil recovery to biological growth~\cite{Pelce}, (ii) remarkable universal asymptotic shapes, it exhibits~\cite{ST,TS,Pelce,KKL,98PRL,TS14}, and (iii) discoveries of deep intriguing connections of LG to quantum gravity~\cite{MWZ} and the quantum Hall effect~\cite{ABWZ}. In {\it mathematics} the Laplacian growth appears so exciting because it possesses beautiful and powerful properties, unusual for most of nonlinear PDEs, such as infinitely many conservation laws~\cite{Rich72} and closed form exact solutions~\cite{Kuf,SB,MS2,AMZ2,TS14}. A new splash of intense activity in LG (see~\cite{MPT} for a review) was provoked by the work~\cite{MWZ}, where strong connections of LG with major integrable hierarchies and the theory of random matrices were established. 

{\it Mathematical formulation} of LG is (deceptively) simple: a droplet of air, $D^+(t)$, where $t$ is time, is surrounded by a viscous fluid, $D^-(t) = \mathbb C/D^+(t)$, called $D(t)$ for simplicity. Both liquids are sandwiched between two parallel close plates. Fluid velocity in $D(t)$ obeys the Darcy law, ${\bf v} =  -\nabla p$ (in scaled units), where $p(z,\bar z)$ is pressure and $z=x+iy$ is a complex coordinate on the plane. Because of incompressibility, $\nabla \cdot {\bf v} = 0$, then $\nabla^2 p = 0$ in $D$, except points with sources, which provide growth. Also, $p=0$ at the interface, $\Gamma(t)=\partial D(t)$, between two fluids, if to neglect surface tension. The kinematic identity requires that normal interface velocity, $V(\xi)$, ($\xi \in \Gamma$), equals to the fluid normal velocity at the interface, thus
\begin{equation}\label{Vint}
V(\xi) = - \partial_n p(\xi),
\end{equation}
where $\partial_n$ is a normal derivative.

{\it Diffusion-limited aggregation} is a process where equal particles are issued {\it one by one} from infinity and diffuse until they stick to a growing cluster~\cite{WS81}.  Remarkably, all grown clusters are monofractals with the numerically obtained Hausdorff dimension $D_h = 1.71 \pm 0.01$, which appears to be  robust and universal~\cite{Halsey} (independent on geometrical details). Analytic derivation of this number remains a long-standing challenge in non-equilibrium physics despite numerous efforts~\cite{Halsey}.  Surprisingly, the same fractal dimension was observed in several Laplacian growth experiments~\cite{Coud,Praud}, where the process is continuous and deterministic.

{\it We have unifed} LG and DLA as two opposite limits  (classical and quantum respectively) of a stochastic Laplacian growth, where instead of one particle the source emits $K \ge 1$ uncorrelated particles per time unit. The DLA, when $K=1$, can be called a quantum limit of this process, as correlations between particles in this case are maximal. The next particle always ``feels'' a slight change of the interface, caused by a previously landed particle, while both would be totally uncorrelated if emitted simultaneously.

By using simple combinatorics we introduce below probability $\mathcal P$ of different growth scenarios and define the action as $\mathcal A = -\hbar\log {\mathcal P}$, where $\hbar$ is the particle area. Then we show that in the limit, $K \to \infty$, the most probable motion of $D(t)$ (the classical point of the action) is deterministic and obeys the Laplacian growth equation. Thus, $K \to \infty$, is the classical limit of this theory.

It is valuable that the action for Laplacian growth comes so simply from growth probability. For it is known that to find a functional, which extremum gives equations for dissipative motion, is much harder, than for frictionless processes, whose Lagrangian or Hamiltonian structure is often straightforward~\footnote{There exists a well-known Martin-Siggia-Rose method~\cite{MSR}  to generate actions for stochastic  equations (which has been never applied to LG).   But we present here very different from~\cite{MSR} and much more direct approach, which reflects all peculiarities of LG and DLA.}.  So far the Laplacian growth equation was derived only as the approximation of viscous hydrodynamics.

{\it Electrostatics and Gibbs-Boltzmann statistics.}  We found that the action for an arbitrary $D(T)$ is fully characterized by harmonic measures, $\mu_n(A_m)$, 
for sources at $A_m$ and by their strengths, $Q_m$.  Then we derived that the action equals a time integral from the entropy, which is LHS of~\eqref{S-E} below. Surprisingly, this purely probabilistic expression can be transformed to a sum of electrostatic potentials created at $a_m$ (the conformal image of $A_m$) by charges induced on the unit circle, kept at zero potential:
\begin{equation}\label{S-E}
\sum_{m,n} Q_m \mu_n(A_m) \, \log \frac{\mu_n(\infty)}{\mu_n(A_m)} = \sum_m Q_m \log(1-|a_m|^2).
\end{equation}
From~\eqref{S-E} it follows that growth probabilities for these patterns obey the usual equilibrium Gibbs-Boltzmann distribution:
\begin{equation}
\mathcal P[D(T)] = \exp\left\{-\frac{1}{K\hbar}\int_0^T dt \sum_m Q_m\log(1-|a_m(t) |^2)\right\},
\end{equation}
where $K\hbar$ serves as temperature. This conclusion opens novel possibilities for analyzing non-equilibrium growth processes by tools of equilibrium statistical physics.

{\it Structure of the paper} is straightforward: after introducing harmonic measure and conformal map we derive the classical Laplacian growth~\eqref{LG} from elementary probability formulae and introduce virtual classical sources, which cause ``non-classical'' complex shapes. This helps to present the growth probability of non-classical shapes in terms of classical sources~\eqref{P-mu}. Then we transform the entropy~\eqref{P-mu} to electrostatic energy~\eqref{P-w1} in the ``conformal'' $w$-plane and~\eqref{Plog1} in the physical $z$-plane. Finally, we reveal physical significance of growth probabilities~\eqref{P-w1} and~\eqref{Plog1}, which are two main results of this work, and establish connections with modern mathematical physics.

{\it The harmonic measure}, $\mu_{D}(\xi_n, A)$, is important in what follows. For simplicity we will skip below the label $D$ and will often refer to $\mu_{D}(\xi_n, A)$ just as $\mu_n(A)$.  Let's partition the boundary, $\Gamma$, into $N \gg 1$ little fragments of the size $|d\xi_n|$, so $n = 1,2,\dotsc, N$. Then the harmonic measure, $\mu_n(A)$, for the $n$-th fragment between $\xi_n$ and $\xi_n + d\xi_n$, with the source at $A$, is defined~\cite{harm.measure} as
\begin{equation}\label{mu-def}
\mu_n(A) = -\frac{\partial_n G_D(\xi_n, A)}{2\pi}\, |d\xi_n|, \qquad \xi_n \in \Gamma,
\end{equation}
where $\partial_n$ is a normal derivative, and $G_{D}(z, \zeta)$ is the Green function of the domain $D$. By definition, $G_{D}(z, \zeta)$ is a harmonic function in $D$, except at $z=\zeta$, where $G(z, \zeta)$ diverges as $\log|z-\zeta|$~\cite{Schiffer}, and also $G_{D}(z, \zeta) = 0$ at the boundary $\Gamma$. In electrostatics $ \mu_n(A)$ is a charge distribution induced at $\xi_n \in \Gamma$  by a unit charge at $A$ to keep $\Gamma$ equipotential. But in this work the harmonic measure is a probability for a Brownian particle, issued at $A \in D(t)$, to land  between $\xi_n$ and $\xi_n +d\xi_n$ at $\Gamma$.

\begin{figure}[h]
\centering
\includegraphics[width=1\columnwidth]{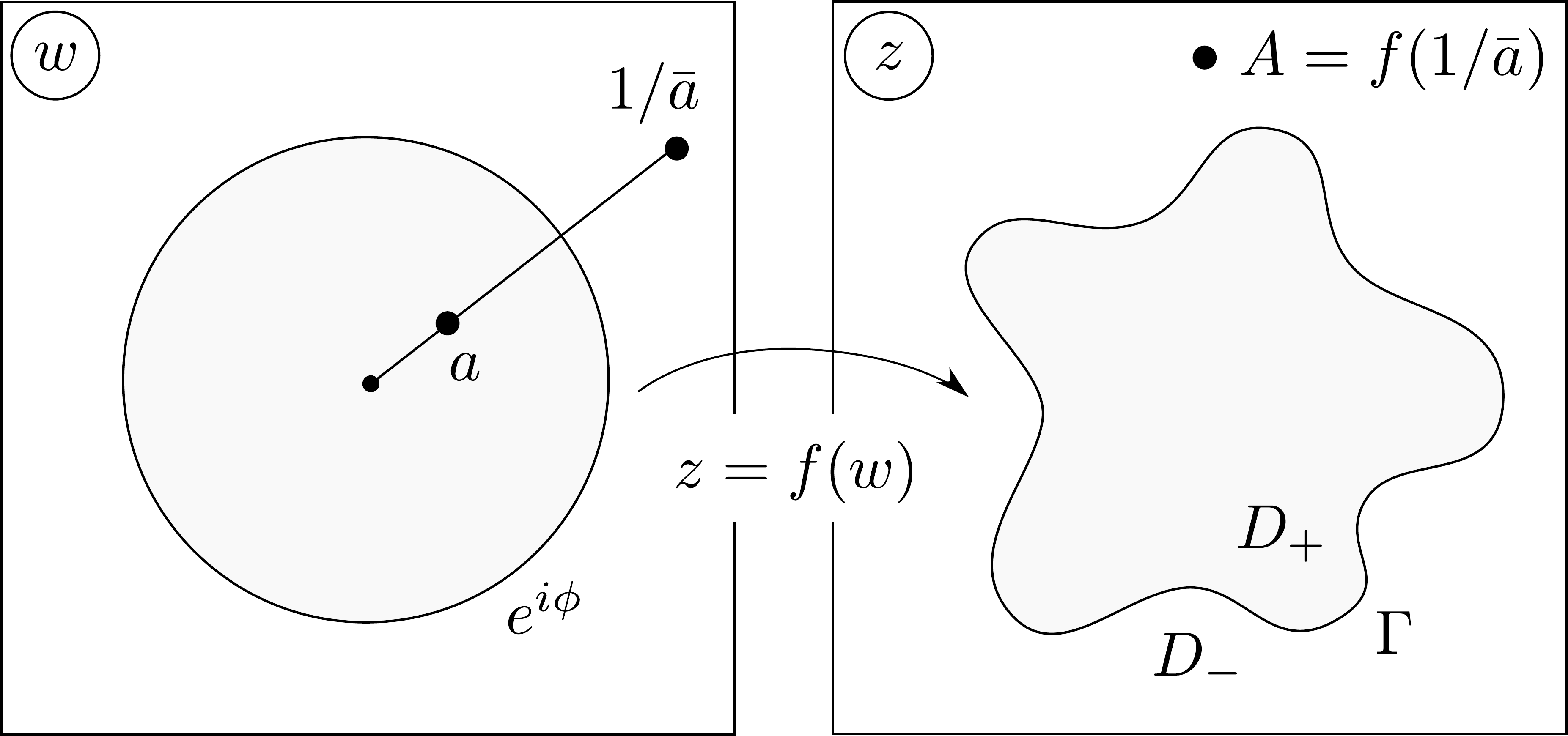}
    \caption{\label{map}Conformal map $z=f(w)$ from the exterior of the unit circle to $D_-$, so that $\infty = f(\infty)$, the conformal radius, $r = f'(\infty) > 0$, and $A = f(1/\bar a)$.}
\end{figure}

{\it Conformal mapping.}  Harmonic nature of $G_{D(t)}(z,A)$ suggests the (time dependent) conformal mapping, $z=f(w)$, from the exterior of the unit circle at the auxiliary complex $w$-plane to the domain $D(t)$ at the physical $z$-plane. Then the unit circle, $w=e^{i\phi}$, maps to $\Gamma(t)$: $\xi = f(e^{i\phi}) \in \Gamma(t)$.  Setting $\infty \to\infty$ and $f'(\infty)>0$ makes this map unique. Each fragment $|d\xi| \in \Gamma$ is mapped from a little arc $d\phi$ lying between $\phi$ and $\phi + d\phi$ at the unit circle, and the source $A = f(1/\bar a)$, where $a$ is a singularity of $f(w)$, and $|a| < 1$ (see Fig.~\ref{map}).

\begin{figure}[h]
\centering
\includegraphics[width=1\columnwidth]{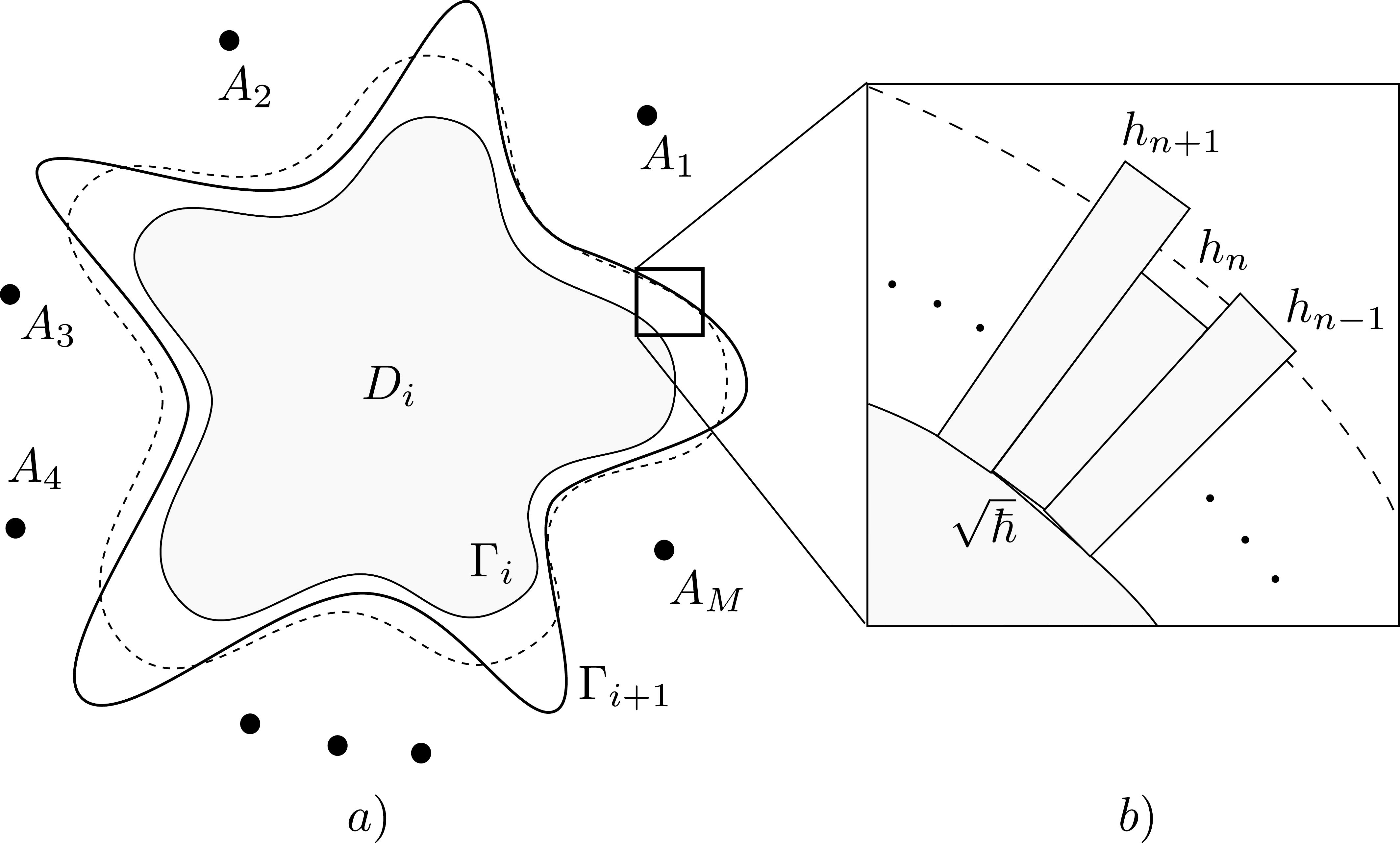}
    \caption{\label{layer}a) Stochastic growth of a single layer, $D_{i+1}/D_i$: Here a thin line is $\Gamma_i = \partial D_i$ formed during the first $i$ time units; a dashed line represents classical (deterministic) LG for a single source at $\infty$ during $(i+1)$-st unit, and a solid line, $\Gamma_{i+1}$ is an external boundary of a stochastic layer, $D_{i+1}/D_i$ grown per elementary time unit, $\delta t$. This stochastic layer is equivalent to a classical layer, grown in the presence of $M$ {\it virtual} sources located at $A_1$, $A_2$,\ldots, $A_M$. b) Three consecutive fragments of $\Gamma_i$, partitioned onto $N \gg1$ equal pieces of the size, $\sqrt \hbar$, after stochastic growth during the $(i+1)$-st time unit. The heights of grown columns equal to $h_{n } = \sqrt \hbar k_{n }$.}
\end{figure}

{\it Calculating $\mu_n(A)$} is simple at $w$-plane, chosen instead of $z$-plane, since $\mu_n( A)$ is conformally invariant.  From the Green function outside the unit circle, $G(w, 1/\bar a) = \log(|1-\bar a w|/|w-a|)$, and~\eqref{mu-def} we obtain
\begin{equation}\label{mu-w}
\mu_n(A)= 
{\cal R}e \frac{e^{i\phi_n} +a}{e^{i\phi_n} - a}\,\frac{d\phi_n}{2\pi}, \ \text{where}\ |d\xi_n| = |f'(e^{i\phi_n})|\, d\phi_n.
\end{equation}

{\it Stochastic Laplacian growth}, \ref{map} unifying LG and DLA as two opposite limits, differs from DLA by $K \geq 1$ (instead of one) of simultaneously issued {\it independent} particles of area~$\hbar$ from several sources. The particles are curvilinear quadrangles formed by equipotential and stream lines, generated by the probability field. A growing domain is initially a unit circle, so $\hbar \ll 1$. The particles quickly diffuse until they stick to a growing cluster (forming on its surface an external layer of the area $K\hbar$) per characteristic time, $\delta t$, which determines a timescale (a unit time) in the problem. Since they diffuse and stick at the interface much faster than the interface grows, $\delta t$ can be treated as the small time interval. The $K$ issued uncorrelated particles are to be distributed into $N$ bins of the boundary with the probabilities, expressed by the harmonic measure~\eqref{mu-def}, such that the particles stuck to the same bin form a column. Also, instead of a single source at $\infty$ there are $M$ uncorrelated sources at $A_m \in D(t)$ ($m=1,2,\dotsc,M$), and $m$-th source emits $K_m$ particles simultaneously. Thus, $\sum_{m=1}^M K_m = K$, a total number of particles (see Fig.~\ref{layer}).

Since these particle sources become fluid sinks with rates $Q_m$ in the continuous (hydrodynamical) limit, the corresponding partial area increments are equal: 
\begin{equation}\label{KQ}
	K_m \hbar = Q_m \delta t.
\end{equation}

{\it A single layer}, grown by $K$ particles issued from $M$ sources at $A_m$, is defined by ${\bf k} = \{k_{m1},\dotsc, k_{mN}\}_{m=1}^M$, where $k_{mn}$ is a number of particles deposited from $A_m$ and landed to the $n$-th fragment. Its probability is given by the multinomial formula, $P({\bf k}) = \prod_{m=1}^M K_m! \prod_{n=1}^N \mu_n(A_m)^{k_{mn}}/k_{mn}!$, which in the Stirling approximation, $K_m \gg 1$, takes a form of the Kullback-Leibler entropy~\cite{KL} (the distance between two distributions): 
\begin{equation}\label{P-Stir}
P({\bf k}) =  \exp\left\{-\sum_{n=1}^N \sum_{m=1}^M k_{mn} \log \frac{k_{mn}}{K_m \mu_n(A_m)}\right\}.
\end{equation}
Derivation of the Laplacian growth equation, (eqs.~\eqref{LG} and~\eqref{LG1} below). Variation of~\eqref{P-Stir} with the constraints, $K_m = \sum_{n=1}^N k_{nm}$, shows that $P$ is maximal when 
\begin{equation}\label{k-cl}
k_{mn}^* = K_m \mu_n( A_m).
\end{equation} 
This maximum is exponentially sharp when $\hbar \to 0$, so all fluctuations around $k_{mn}^*$ are suppressed.  Hence $k_{mn}^*$ is a classical trajectory for this stochastic process. It describes the deterministic Laplacian growth with $M$ sources, $Q_m$ at $A_m = f(1/\bar a_m)$. Indeed, because the normal displacement at $\xi_n$ is $V(\xi_n) \delta t = \sqrt \hbar \sum_m k^*_{mn}$ (the particles are approximate squares, $\sqrt \hbar \times \sqrt \hbar$), it readily follows from~\eqref{mu-w},~\eqref{KQ},~\eqref{k-cl}, and the identity, $V(\xi) = {\cal I}m(\bar f_t f_\phi)/|f_\phi|$, that
\begin{equation}\label{LG}
{\cal I}m (\bar f_t f_\phi) = \sum_{m=1}^M \frac{Q_m}{2\pi} {\cal R}e \frac{e^{i\phi} +a_m(t)}{e^{i\phi} - a_m(t)},
\end{equation}
which is the classical Laplacian growth with $M$ sources.  If $M=1$, $Q_1(t) = Q$, and $A_1 = \infty$,~\eqref{LG} takes a form 
\begin{equation}\label{LG1}
{\cal I}m (\bar f_t f_\phi) = Q/(2\pi), 
\end{equation}
which was intensely studied earlier~\cite{GP-K,Pelce,MWZ,MPT,Gillow,GTV}  and addressed by the stochastic LG in~\cite{Gruzberg}. 

Thus, it turned out possible to derive the Laplacian growth equation directly from variational calculus based on elementary combinatorics.  So far this equation was possible to deduce only from viscous hydrodynamics or kinetics~\cite{Pelce}.

{\it Growth probability as the sum over scenarios.}  If growth continues until time $T\gg\delta t$ ($T/\delta t$ is integer), then different scenarios to arrive from $D(0)$ to $D(T)$ exist, depending on ordering operation of the sources $A_m$ in time. The total probability is the sum over probabilities of all scenarios, ${\cal P}_{\rm total}[D(T)] =  \sum_{({\bf \Bbbk})} {\cal P}(\bf \Bbbk)$, where ${\bf \Bbbk}=\{{\bf k}_i\}_{i=1}^{T/\delta t}$ labels different scenarios, and ${\bf k}_i$ denotes the $i$-th layer. Probability of a single scenario equals
\begin{equation}\label{P-sc}
{\cal P}({\bf \Bbbk}) = \prod_{i=1}^{T/\delta t} P({\bf k}_i),
\end{equation}
where $P({\bf k}_{i+1})$ is a {\it conditional} probability for the layer ${\bf k}_{i+1}$ to grow over the domain $D_i\equiv D(i \delta t)$. Since $P({\bf k}_{i+1})$ is independent of prehistory, $t < i\delta t$, the product in~\eqref{P-sc} is a Markovian chain, and the sum over all scenarios extends standard path-integration to scenarios  in the space $\{\Gamma(t)\} \times [0,T)$. Taking $P({\bf k}_i)$ from~\eqref{P-Stir}, $K_m$ from~\eqref{KQ}, and transforming~\eqref{P-sc}, we obtain, when $\delta t\to0$, that $\mathcal P({\bf \Bbbk}) = \exp\{-\mathcal A({\bf \Bbbk})/\hbar\}$, where we defined the stochastic action ${\cal A}({\bf \Bbbk})$ as
\begin{equation}\label{A-def}
{\cal A}({\bf \Bbbk}) = \int_0^T \frac{Qdt}{K} \sum_{m,n=1}^{M,N} k_{mn}(t) \log \frac{k_{mn}(t)}{k^*_{mn}(t)}, \ (Q=\sum_m Q_m).
\end{equation}
For a single source at infinity, $M=1$ and $A_1 = \infty$, the action~\eqref{A-def} 
takes a form
\begin{equation}\label{A0}
{\cal A}_0({\bf \Bbbk}) = \int_0^T \frac{Qdt}{K} \sum_{n=1}^{N} k_{n}(t) \log \frac{k_{n}(t)}{K \mu_n(\infty)},
\end{equation}
where we replaced $k^*_{1n}$ by its classical value~\eqref{k-cl}.

{\it Virtual sources}.  We see from~\eqref{LG1} that in LG with a single source at infinity an initial circle, $z=r_0 e^{i\phi}$, stays as a circle, $z=r(t) e^{i\phi}$.  It is the classical trajectory of the stochastic action~\eqref{A0}.  But in experiments with a single source far enough from $\Gamma(t)$ to be treated as at infinity~\cite{ST, Coud, Praud}, complex irregular interfaces, caused by intrinsic instabilities of the process, are always observed. 

Remarkably, any non-circular domain $D(T)$ bears ``fingerprints'' (singularities of the Schwarz function \footnote{The Schwarz function of the curve, $\bar z = {\cal S}(z)$, when $z$ belongs to the curve, is described in~\cite{Davis}}, $\mathcal S(T,z)$ for $\Gamma(T)$, lying in $D(T)$) left by sources operated at earlier times, $t < T$. These complex shapes, $D(T)$, correspond to non-classical trajectories of the stochastic action~\eqref{A0}. Thus, all deviations of  ``non-classical'' $D(T)$ from a growing circle we attribute to these {\it virtual} sources at $A_m$, working in their {\it classical} regime \footnote{Earlier explanations of complex shapes by tiny initial irregularities~\cite{MS2,AMZ2} contradict to the fact that, as a rule, unstable nonlinear systems quickly forget initial details. Thus the presented virtual random sources approach is more physical to explain fronts complexity.}. This process is described by the classical equation~\eqref{LG}, but with {\it time-dependent} virtual sources, $Q_m$ (except $Q_1$ at $\infty$), which cause observable deflections of $\Gamma(t)$ from the classical path, prescribed by~\eqref{LG1} \footnote{Little factors, neglected in~\eqref{LG}, such as surface tension, boundary effects, and kinetics, certainly contribute to shape deviation from~\eqref{LG1}, but being coarse-grained limits of different realizations of virtual sources at $A_m(t)$, these factors can be included in the stochastic treatment here.}. Each virtual source $A_m$ contributes to the growth probability by the action~\eqref{A0} with $k_n$ given by~\eqref{k-cl}. By summing up contributions of  the independent virtual sources during $\delta t$ we obtain the logarithm of the probability of the non-classical layer:
\begin{equation}\label{P-mu}
\log P({\bf k}_i) = \sum_{m,n = 1}^{M,N} K_{mi} \mu_{n,i-1}(A_m) \log \frac{K \mu_{n,i-1}(\infty)}{K_{mi} \mu_{n,i-1}(A_m)},
\end{equation}
where $\mu_{n,i}(A_m)$ is referred to $D_i$. This is the Kullback-Leibler entropy mentioned above.

{\it Contribution of  $\log(K/K_{mi})$} in~\eqref{P-mu} equals $K\log K - \sum_{m=1}^M K_{mi} \log K_{mi} = \log\{K!/\prod_{m=1}^M K_{mi}!\} = \log {\cal N}_i$ when $K_m \gg 1$. Here ${\cal N}_i$ is the number of partitions of $K$ particles into $M_i$ groups (sources), $K_1, \dotsc, K_{M_i}$, at $i$-th time step. Replacing $\mu_{n,i}(A_m)$ by $-\,\partial_n G_i(\xi, A_m)\, |d\xi|/(2\pi)$ under the logarithm in~\eqref{P-mu} we rewrite~\eqref{P-mu} in the continuous limit, $|d\xi_n| \to 0$, as
\begin{equation}\label{P-G}
\log \frac{P({\bf k}_{i+1})}{{\cal N}_{i+1}} = \sum_{m=1}^{M_{i+1}} K_{m,{i+1}} \oint_{\Gamma_{i}} \mu_i(\xi, A_m) \log \frac{\partial_n G_i(\xi,\infty)}{\partial_n G_i(\xi, A_m)},
\end{equation}
where $\Gamma_i=\partial D_i$ and $G_i = G_{D(i\delta t)}$.

{\it A Dirichlet problem of recovery} harmonic functions by their boundary values is of great help in calculating this integral. Notice that $\partial_n G(\xi,\infty) = -|w'(\xi)|$, where $w = w(z)$ is inverse to the conformal map, $z=f(w)$, introduced earlier. Since $\log |w'(z)|$ is harmonic in $D$, then $\log |w'(\xi)|$ is the boundary value of this harmonic function, so the contribution of the numerator to the integral in~\eqref{P-G} equals
\begin{equation}\label{Measure}
\oint_{\Gamma}\mu(\xi,A) \log|\partial_n G(\xi,\infty)| = \log |w'(A)|.
\end{equation}
(The label $i$ is omitted in~\eqref{Measure}-\eqref{Gint} as unnecessary.)

{\it Key observation.} Contribution of the denominator into the integral~\eqref{P-G} can be rewritten in a remarkably simple way. Presenting $G(z,A) = {\cal R}e \,W(z,A)$ as $G(z,A) = G^+(z,A) + G^-(z,A) = {\cal R}e \, (W^+ + W^- )$, where $G^+ = {\cal R}e W^+ = \log|z-A|$, we obtain from $-\partial_n G = |\partial_z W|$ that
\begin{equation}\label{logG}
\log |\partial_n G(\xi, A) | =   {\cal R}e \, \log\left(\frac{1+ (\xi-A)\partial_\xi W^-(\xi, A)}{\xi-A}\right).
\end{equation}
Since near infinity $\partial_z W^-(\xi, A) = - 1/\xi + {\cal O}(\xi^{-2})$, then subtracting $G(\xi, \infty) = 0$ at $\Gamma$ from~\eqref{logG} makes the expression, $\log|1+ (\xi-A)\,\partial_z W^-(\xi, A)| - G(\xi, \infty)$, harmonic for $\xi$ everywhere in $D$.  Therefore the corresponding integral in~\eqref{P-G}, being a solution of the Dirichlet boundary problem, is the difference between $\oint_{\Gamma} \mu(\xi,A)\log|\xi-A|=  -G^-(A, A) + G(A, \infty)$ from the denominator and $\oint_{\Gamma} \mu(\xi, A)\log|1+ (\xi-A)\partial_\xi W^-(\xi, A)| = - G(A, \infty)$ from the numerator of~\eqref{logG}. After adding~\eqref{Measure} to this difference we finally obtain the remarkable identity,
\begin{multline}\label{logG3}
\oint_{\Gamma} \mu(\xi, A) \log \frac{\partial_n G(\xi,A)}{\partial_n G(\xi,\infty)} = \\ = 
G^-(A, A) - 2G(A, \infty) - \log |w'(A)|,
\end{multline}
which has clear electrostatic interpretation shown below.

{\it In $w$-plane} the integral~\eqref{logG3} can be further simplified. Namely, because of~\eqref{mu-w}, the integral~\eqref{logG3} is easily calculated to equal
\begin{equation}\label{Gint}
\oint_{|w| = 1} \frac{1-|a|^2}{|w-a|^2} \log  \frac{1- |a|^2}{|w - a|^2}\frac{dw}{2\pi i w} 	= - \log\left( 1-|a|^2 \right),
\end{equation}
where $A = f(1/\bar a)$, as said above. This is the Robin function~\cite{Gust}, which is a potential at $a$ created by charges induced by a unit charge at $a$ on the unit circle, kept at zero potential. Thus, the entropy~\eqref{P-mu} was transformed to electrostatic energy, and with help of~\eqref{Gint} the probability~\eqref{P-G} for a single layer can be compactly rewritten in a form of the Gibbs-Boltzmann distribution, (implying that probability $P_i = C\exp(-\beta E_i)$, where $E_i$ is the energy of $i$-th state and $\beta$ is a positve constant):
\begin{equation}\label{P-w}
	P({\bf k}_i) = \mathcal N_i\exp\left\{\sum_{m=1}^{M_i} K_m\,\log(1-|a_m|^2)\right\}.
\end{equation}
Then in the limit $\delta t \to 0$ the scenario probability~\eqref{P-sc} becomes (after defining ${\cal N} = \prod_{i=1}^{T/\delta t} {\cal N}_i$)
\begin{equation}\label{P-w1}
\mathcal P ({\Bbbk})=  \mathcal N \exp\left\{\int_0^T \frac{dt}{\hbar} \sum_{m=1}^{M(t)} Q_m \log(1-|a_m(t)|^2)\right\},
\end{equation}
and $A_m = f(1/\bar a_m)$ provides time-dependence of~$a_m$.

{\it In $z$-plane} the integral~\eqref{logG3} allows to recast~\eqref{P-G} in another remarkable way: adding~\eqref{P-G} over all $i$ and assuming one source per one time unit (we set $m=i$ for convenience) we obtain via~\eqref{P-sc} the logarithm of probability of the given scenario, 
\begin{equation}\label{GlogG}
	\log \frac{{\cal P}({\bf \Bbbk})}{{\cal N}} 
= - \sum_{i=1}^{T/{\delta t}} K_i \oint_{\Gamma_{i-1}} \mu(\xi,A_{i})
\log \frac{\partial_n G(\xi, A_{i})}{\partial_n G(\xi,\infty)}.
\end{equation}
Contribution from the denominator in the RHS of~\eqref{GlogG} equals to the following neat expression, which depends on the final domain $D(T)$ only, but not on a particular way to arrive to it~\cite{OM}:
\begin{equation}\label{logM}
\log{\cal M}_{D(T)} = \frac{1}{2\hbar}\left\{{\rm Area} \,D(T) + \oint_{\Gamma(T)} \log |w_T'(\xi)| \, \frac{\bar \xi d\xi}{i}\right\}.
\end{equation}
As to the numerator in~\eqref{GlogG}, it is shown~\cite{OM} to equal
\begin{equation}\label{I}
	I = - \oint_{\Gamma_{i-1}} \mu(\xi, A_i) \oint_{\Gamma_{i-1}} \mu(\eta,A_i) \log|\xi-\eta|,
\end{equation}
which is the energy of self-interacting charge induced on $\Gamma_{i-1}$ with density $\mu(\xi,A_i)$.
After transforming contour integrals over $\Gamma_{i-1}$ into integrals over the layer, $l_i = D_{i}/D_{i-1}$, it becomes (see~\cite{OM})
\begin{equation}\label{I1}
(Q_i\delta t)^2 I = - \int_{l_i}\int_{l_i} \log|z-\zeta|\,d^2z\,d^2\zeta + \pi\int_{l_i} {\cal A}_{i}(z)\,d^2z,
\end{equation}
where ${\cal A}_i(z) = |z|^2/2 - {\cal R}e \int_0^z {\cal S}(i\delta t, z') \,dz'$ is the so-called modified Schwarz potential~\cite{DKh}. Then the probability of the whole scenario~\eqref{P-sc} equals
\begin{multline}\label{Plog1}
	\mathcal P({\bf \Bbbk})=\mathcal N \mathcal M_{D(T)}\prod_{i=1}^{T/\delta t} \exp\left\{\frac{1}{K_i\hbar^2}\times\right.
\\
\times
\left.\left(\int_{l_i}\int_{l_i} \log|z-\zeta|\,d^2z\,d^2\zeta  - \pi\int_{l_i} {\cal A}_i(z)\,d^2z\right)\right\}.
\end{multline}
Strikingly, despite {\it non-equilibrium} nature of LG, in both $w$ and $z$ planes, the growth probabilities~\eqref{P-w1} and~\eqref{Plog1}, appear in a form of the {\it equilibrium} Gibbs-Boltzmann distribution, since exponents in these formulae are electrostatic energies (up to multiplicative constants).

Finally, interesting connections were found between the stochastic LG and modern mathematical physics.

{\it Connection to the Liouville theory.} Remarkably, the Robin function, $G^-(w,w)= - \log(1-|w|^2)$, obtained in~\eqref{Gint}, obeys the (integrable) Liouville equation, vital for the theory of non-critical strings~\cite{Polyakov}, 
\begin{equation}\label{G-Lio}
\nabla_w^2 G^-(w,w) = 4 \exp\{2 G^-(w,w)\}.
\end{equation}
Thus, the probability~\eqref{P-w} appears to be the classical limit of the multipoint correlation function of ``light'' exponential operators in the Liouville theory on the pseudosphere~\cite{ZZ} (see~\cite{OM} for details).

{\it Connection to the tau-function for the Toda hierarchy.} The main result~\eqref{Plog1} for a scenario probability can easily be rewritten through the following integrals 
\begin{equation}\label{tau-l} 
f_i = -\frac{1}{\pi^2} \int_{l_i} \int_{l_i} \log\left|\frac{1}{z}-\frac{1}{\zeta}\right|\,d^2z\, d^2\zeta,
\end{equation}
which coincide with the tau-function for analytic curves~\cite{KKMWZ} after replacing a layer $l_i$ by a simply connected finite domain $D$. The obtained expression (skipped here for want of space) connects growth probability to the dispersionless 2D integrable Toda hierarchy.

Replacement of $l_i$ by $D$ in~\eqref{tau-l} implies interaction of all layers constituting the domain. However, in our case the layers, $l_i$, enter~\eqref{Plog1} additively, and so are independent. Hence the sources, participated in growth of different layers, are also mutually independent (but not commutative, contrary to the classical deterministic LG).

{\it Summarizing}, we emphasize the derivation of the Laplacian growth equation~\eqref{LG} from the action functional using elementary combinatorics, and unexpected relation between the entropy~\eqref{P-mu} and electrostatic energies on the $w$-plane~\eqref{Gint} and the $z$-plane~\eqref{I}. As the result, the growth probabilities satisfy the Gibbs-Boltzmann statistics, suggesting applications of weakly non-equilibrium thermodynamics to this highly unstable and non-equilibrium process.

{\it In conclusion,} we state the expected impact of the results, obtained in this paper,

- {\it for LG and DLA}: The presented theory promises to elucidate derivation of the DLA fractal spectrum and unexplained selection problems in LG.

- {\it for non-equilibirum physics}: It appears possible now to address highly non-equlibrium growth in the framework of {\it linear} non-equilibrium thermodynamics~\cite{OS} and to reinterpret complex pattern formation as a self-organizing non-equilibrium process.

- {\it for other branches of physics}: Remarkably, the growth probability~\eqref{Plog1} links stochastic LG with a growth of an electronic droplet in a quantum Hall effect~\cite{ABWZ}. It allows us to reformulate our model in terms of normal random matrices~\cite{NRM},  which underly 2$D$ quantum gravity~\cite{GR}.

{\it The next step} is to go beyond the classical limit, which was the subject of this work, and to study {\it quantum stochastic} Laplacian growth, where the correlations between particles become important, i.e. when $K$ is small.

\end{document}